\documentclass[aps,prl,reprint,superscriptaddress]{revtex4-1}

\usepackage[version=3]{mhchem}
\usepackage{bm}
\usepackage{amsmath}
\usepackage{amssymb}
\usepackage{color}
\usepackage{epsfig}
\usepackage{hyperref}
\usepackage[section]{placeins}

\newcommand{\nix}[1]{}

\renewcommand{\epsilon}{\varepsilon}
\renewcommand{\phi}{\varphi}

\begin{document}

\title{Revisiting the physics of Fano resonances for nanoparticle oligomers}

\author{Ben Hopkins}
\affiliation{Nonlinear Physics Centre, Australian National University, Canberra, ACT 2602, Australia}
\email{ben.hopkins@anu.edu.au}

\author{Alexander N. Poddubny}
\affiliation{Ioffe Physical-Technical Institute of the Russian Academy of Sciences, St. Petersburg 194021, Russia}
\affiliation{National Research University for Information Technology, Mechanics and Optics, St. Petersburg 197101, Russia}

\author{Andrey E. Miroshnichenko}
\affiliation{Nonlinear Physics Centre, Australian National University, Canberra, ACT 2602, Australia}

\author{Yuri S. Kivshar}
\affiliation{Nonlinear Physics Centre, Australian National University, Canberra, ACT 2602, Australia}
\affiliation{National Research University for Information Technology, Mechanics and Optics, St. Petersburg 197101, Russia}

\begin{abstract}
We present a new and robust approach for interpreting the physics of Fano resonances in planar oligomer structures of both metallic and 
dielectric nanoparticles.  We reveal a key mechanism for Fano resonances by demonstrating that such resonances can be generated purely from the interference of 
nonorthogonal collective eigenmodes, which are clearly identified based on the coupled-dipole approximation. 
We prove analytically a general theorem to identify the number of collective eigenmodes that can be excited in 
ring-type nanoparticle oligomers and further demonstrate that no dark mode excitation is necessary for existence of 
Fano resonances in symmetric oligomers.
As a consequence, we unify the understanding of Fano resonances for both plasmonic and all-dielectric oligomers.
\end{abstract}

\maketitle

\section{Introduction}

Recently a lot of  attention was attracted to {the physics of} Fano resonances\cite{Fano1961} in  nanoscale oligomer structures composed of plasmonic nanoparticles.  The current understanding of Fano resonances  in these symmetric oligomers relies on an interference of super- and sub-radiant collective modes, such as those created from the interaction between a symmetric ring of nanoparticles and a single central nanoparticle\cite{Miroshnichenko2010, Lukyanchuk2010, Davis2010}.  With very few exceptions\cite{Frimmer2012, Lovera2013}, this interference is described for specific plasmonic oligomers, where a directly-excited super-radiant mode interferes destructively with  an indirectly-excited `dark mode' (or trapped mode).  {By `dark mode' we refer to any mode that cannot  couple directly with an incident plane wave. More formally, when given an oligomer's symmetry group, if the irreducible representation of a particular mode is mutually exclusive to the irreducible (or reducible) representation of the incident field, then that mode cannot be excited by the incident field and it is defined as dark~\cite{Sonnefraud2010}.  However we use a common extension to this definition and also consider dark modes as being modes whose projection onto the incident field is zero.  Thus, such dark modes cannot couple directly to the incident field, although they may transform according to the same irreducible representation as the incident field.  The current understanding is that the latter form of dark modes can still be excited in symmetric oligomers through the coupling to a bright plasmonic mode via near-field hybridization~\cite{Prodan2003, Prodan2004, Nordlander2004, Brandl2006, Bao2010, Mirin2009, Chuntonov2011,Christ2008, Fan2010, Lukyanchuk2010, Thyagarajan2013, Giannini2011, Agio2013}}
While the concept of dark modes inducing Fano resonances in oligomer  structures has been discussed in  many  {studies}~\cite{ Hentschel2010, Liu2012, Hentschel2011, Artar2011}, the associated definition of the system's modes is deduced from molecular analogues that { {\em do not fully resemble the electromagnetic coupling in oligomer systems}}.  For this reason, such modes do not represent the real eigenmodes of the oligomer system.  This observation has led to the necessary distinction between bare (approximate) and dressed (real) modes in the existing literature when regarding plasmon hybridization from the perspective of quantum optics ~\cite{Frimmer2012, Barnett2002, Hatef2012, Chuntonov2011}.  

Recently, it was predicted  that  Fano resonances should  also occur in all-dielectric symmetric oligomers \cite{Miroshnichenko2012} despite the absence of the necessary hybridization of modes required to excite dark modes. As such, this development calls for {the study of different mechanisms that also lead to Fano type interference and resonances}. 

By revisiting the mechanisms that underpin the general optical properties of oligomer structures, {in this paper} we present a general symmetry theory for analysing their electric and magnetic responses. We show that  generic oligomer structures can be characterised entirely in terms of distinct {{\em electric and magnetic collective eigenmodes}}. In this approach,  we are able to unify both plasmonic and all-dielectric oligomers.  We are then able to build upon work by Frimmer et al.~\cite{Frimmer2012} to  show conclusively that Fano resonances created in  symmetric oligomers can be explained entirely in terms of  interference of these eigenmodes. 
Importantly, this theory does not rely on the hybridization of modes and any mode can only be reliably excited if it couples with {an} incident field. {Therefore,} only bright eigenmodes will be excited except in accidental cases. To demonstrate this {finding,} we  revisit experimental results {that demonstrated} Fano resonances in  gold heptamers and show  that they occur purely due to the interference of the system's bright eigenmodes and do not involve dark mode interference with which they were previously attributed.  

We are further able to address  the types of oligomer systems that can support Fano resonances on the basis that two or more nondegenerate eigenmodes must be excited to permit an interference.  We prove that the response of any symmetric ring-type oligomer consisting of arbitrarily many nanoparticles will be described by precisely {\em two nondegenerate} eigenmodes.  Subsequently, we are able to conclude and numerically demonstrate that ring-type oligomers can support Fano resonances as well.  This idea seems to be in contradiction with current understanding of the origin of  Fano resonances in oligomers where it is typically shown that the removal of a central particle should destroy the resonance~\cite{Hentschel2010, Francescato2011}.
The common practice of adding a central particle to an oligomer system is then re-evaluated as simply increasing the number of (excitable) nondegenerate eigenmodes in the {nanoparticle} system. 

{To amplify the full implications of this eigenmode theory we also consider all-dielectric oligomers where the individual constituent particles have nontrivial electric {\em and} magnetic responses~\cite{Evlyukhin2010, Garcia-Etxarri2011, Kuznetsov2012,  EvlyukhinNL2012, Fu2013}.  We  provide formal and rigorous definitions of the distinct electric and magnetic eigenmodes present in these systems.  As such, we are able to precisely define and justify the existence of purely {\em magnetic Fano resonances} in terms of the new physics discussed in this paper. }
Subsequently, there is immediate relevance of this theory to the emerging field of all-dielectric nanophotonics as it provides {an explicit}  theoretical framework for analyzing both electric and magnetic responses in oligomers.  However, it is also worth acknowledging that the generality of our approach makes it important to broader fields including molecular optics and antenna design.

\section{Eigenmode analysis of oligomers}
The modes of plasmonic oligomer structures are typically understood based on their electric dipole moments  or surface charge distributions~\cite{Fan2010, Agio2013, Chuntonov2011} and, therefore, the analysis of these modes can be carried out from the perspective of the coupled-dipole approximation~\cite{Mulholland1994, Purcell1973, Draine:94}.   In this approximation, the incident field  $\left | E_0\right \rangle$ can be linked to the induced dipole moments $\left |p \right \rangle$ using an interaction matrix $\hat{M}$.
\begin{align}
\left | E_0\right \rangle = \hat M \left |p \right \rangle\;.  \label{eq:dipole equation}
\end{align}
Here the use of the `bra-ket' notation for the incident field and dipole moments is simply to emphasize that each state is the concatenation of all the dipole-moment vectors in the given system.    
In the following, we  derive constraints for a general oligomer's basis vectors and then eigenmodes  using the group theory.  For this regard, it has previously been shown that the interaction matrix must transform according to the symmetry group of the corresponding system's geometry~\cite{Hopkins2013}. The particular symmetry group of the oligomers we are interested in is $D_{Nh}$, where is $N$ is arbitrary.  
 Therefore, the components of a normally incident electric field $E_{0,x}$ and $E_{0,y}$  transform according to the irreducible representation $ E\equiv E_{1}'$ ($E_{1u}$) for odd  (even) $N$~\cite{Dresselhaus2008}. Then we  consider the in-plane components of the electric dipole moments as transforming according to some reducible representation $ P$. Formally, the number of nondegenerate eigenmodes interacting with the electric field is given by the number of times the irreducible representation $E$ is contained in $P$~\cite{Dresselhaus2008}.  From this, we can deduce the dimension of the space that spans all responses of the system to such an incident field, given that each associated eigenmode is doubly-degenerate because $E$ has a rank of two.

Let's consider a plane wave that is normally incident on an oligomer that contains a symmetric ring of $N$ particles, the particles of which are located at the points
\begin{subequations}\begin{align}
 \mathbf r_i=-r\sin\phi_i{\mathbf e_x}+
r\cos\phi_i{\mathbf e_y}\;,\\ \phi_i=\frac{2\pi}{N}(i-1),\quad i=1\ldots N\;. 
\end{align}\end{subequations}
The basis vectors for the response of this ring can then be determined by using the projection operators technique~\cite{McWeeny1963,Dresselhaus2008}. Moreover, we  apply all projection operators to some initial state $\left | e \right \rangle$ in order to produce different linear combinations of all the basis vectors associated with the irreducible representation $E$ and subsequently deduce the dimension of space of the corresponding responses.
 Now we note that a dipole moment at the location of one particular particle is moved through every location on the ring by symmetric rotations.  As such, an initial state that has a dipole moment at only one location on the ring must be linearly independent from all its symmetrically-rotated versions.  By extrapolation, two such initial states that occupy the one location and are linearly independent will, in conjunction with their respective symmetry-rotated versions, form a basis for a $2N$-dimensional space.  {However,} we are only considering responses that transform according the $E$ irreducible representation and are thereby neglecting the $z$-direction, hence the space of responses cannot have more than $2N$ dimensions.  In other words, any response of the system that transforms according to $E$ will be spanned by the projections of these two initial states.  
 To this end we can define the initial state $\left | e \right \rangle$ to have only one nonzero dipole moment at the first particle, and then apply the projection operators~\cite{nussbaum1968} to {obtain} the subsequent polarizations of the other dipoles.  For the symmetry operations $g$ and their matrix representations $\hat{D}(g)$, the projection operator is defined as
\begin{equation}
  \mathcal P^{(\mathrm E)}_{\mu\nu}=\sum\limits_{g}D^{(\mathrm E)}_{\mu\nu}(g)g\;, \label{eq:project}
\end{equation}
where we neglect normalization.  
The idea is that the operator $\mathcal P$ acting upon any state $\left | p \right \rangle$ will yield either a zero or a linear combination of basis vectors that transform according to the vector representation $E$. Hence, we can start with the formal definition of our test state
\begin{equation}
\left | e \right \rangle \equiv (\bm e ,\bm 0,... \bm0)\:, \quad \mathrm{where}\;\;\bm e = (e_{x},e_{y},0)\;.
\end{equation}
Here $e_x$ and $e_y$ are arbitrary in order to produce two linearly independent initial states by varying the polarization of the vector $\bm e$.  All possible basis vectors can then be deduced by varying the indices $\mu$ and $\nu$ in $\mathcal P^{(\mathrm E)}_{\mu\nu}$.  For the following we also introduce the notation to extract the $i^{\mathrm{th}}$ dipole moment from a state
\begin{equation}
 [\left | p \right \rangle ]_{i} \equiv p_{i,x}{\bm e_x}+ p_{i,y}{\bm e_y},\qquad i=1..N\;.
\end{equation}
{We} have chosen to neglect the $z$-direction given that there is no operation in $E$ that  transforms an $x$- or $y$-polarized dipole moment onto the $z$-direction.  Now, because there is only a limited set of symmetry operators that will transform the first particle's dipole moment onto that of the $i^{\mathrm{th}}$ particle, it can be checked using the full nonlocal definitions~\cite{Hopkins2013} of each $g$ that
\begin{subequations}\label{eq:project2}
\begin{align}
 [\mathcal P_{\mu\nu}\left | e \right \rangle]_{i}&=\mathcal P_{\mu\nu, i}\bm e \;, \\
\mathcal  P_{\mu\nu, i}&=
[\hat{D}(R_i)\hat{D}(\sigma_v)]_{\mu\nu} R_i \sigma_v+
\hat{D}(R_{i})_{\mu\nu} R_i \;,
\end{align}
\end{subequations}
where $R_i$ is a  local rotation by the angle $\phi_i$, $\sigma_{v}$ is a local reflection about the $yz$-plane and their matrix representations are
\begin{equation}
  \hat{D}(R_i)=\begin{pmatrix}
            \cos \phi_{i}&-\sin \phi_{i}\\\sin\phi_{i}&\cos\phi_{i}
           \end{pmatrix}\:,\quad \hat{D}(\sigma_v)=\begin{pmatrix}
                                           -1&0\\0&1
                                          \end{pmatrix}\;.
\end{equation}
Noticeably in neglecting the $z$-direction we have not considered the $R_i \sigma_h$ (also known as $S_i$) or $R_i C_2$ operations, however these will be desribed by {exactly the} same symmetry operations and matrix representations as  $R_i$ and  $R_i \sigma_v$ (respectively) for $E$.  For this reason, these extra terms can be neglected by normalization.  
In summary, Eq.~(\ref{eq:project2}) is just a way to write the components of the operated states obtained by acting with $\mathcal P_{\mu \nu}$ upon $\left | e \right \rangle$.
{However,} it allows us to {obtain} explicit forms for the matrix $\mathcal  P_{\mu\nu, i}$
\begin{subequations} \label{eq:original_basis_vectors}
\begin{align}
 \mathcal P_{11,i}=\left(\begin{smallmatrix}
           2\cos^2\phi_i&0\\2\cos\phi_i\sin\phi_i&0
          \end{smallmatrix}\right),&\quad
\mathcal P_{12,i}=\left(\begin{smallmatrix}
           0&2\sin^2\phi_i\\0&-2\cos\phi_i\sin\phi_i
          \end{smallmatrix}\right),\\
\mathcal P_{21,i}=\left(\begin{smallmatrix}
           2\cos\phi_i\sin\phi_i&0\\2\sin^2\phi_i&0
          \end{smallmatrix}\right),&\quad
\mathcal P_{22,i}=\left(\begin{smallmatrix}
           0&-2\cos\phi_i\sin\phi_i\\0&2\cos^2\phi_i
          \end{smallmatrix}\right)\;.
\end{align} \end{subequations}
Evaluating Eq.~(\ref{eq:project2}) then yields only 4 nonzero combinations
\begin{subequations}\label{eq:modes1}
\begin{align}
\mathcal P_{11,i}{\bm e_x}=\left(\begin{smallmatrix}
                                    2\cos^2\phi_i\\2\cos\phi_i\sin\phi_i
                                   \end{smallmatrix}\right),&\quad 
\mathcal P_{21,i}{\bm e_x}=\left(\begin{smallmatrix}
                                    2\cos\phi_i\sin\phi_i\\2\sin^2\phi_i
                                   \end{smallmatrix}\right),\\
\mathcal P_{12,i}{\bm e_y}=\left(\begin{smallmatrix}
                                    2\sin^2\phi_i\\-2\cos\phi_i\sin\phi_i
                                   \end{smallmatrix}\right),&\quad
\mathcal P_{22,i}{\bm e_y}= \left(\begin{smallmatrix}
                                    -2\cos\phi_i\sin\phi_i\\2\cos^2\phi_i
                                   \end{smallmatrix}\right)\;.
\end{align} \end{subequations}
In this {sense,} we have shown that these four basis vectors span all responses of the system that transform according to the irreducible representation $E$.  For later {reference,} it is also worth noting that this approach will also hold for a magnetic incident field and the associated responses, where $E \equiv \mathrm{E}''_1$ ($\mathrm{E}_{1g}$) for even (odd) $N$.  This would be done by simply substituting in a negative version of $\sigma_v$ (and $\sigma_h$) and following the exact same argument as has just been presented here.   
In any case, we have the freedom to create a more intuitive set of basis vectors from linear combinations of the basis vectors presented in Eq.~(\ref{eq:original_basis_vectors})~\cite{Poddubny2011wood}
\begin{subequations} \label{eq:basis mode definitions}
\begin{align}
 \bm p^{(1x)}_i&=\frac1{2\sqrt{N}}(\mathcal P_{11,i}{\bm e_x}+\mathcal P_{12,i}{\bm e_y})=\frac{{\bm e_x}}{\sqrt{N}}\;,\\
\bm p^{(1y)}_i&=\frac1{2\sqrt{N}}(\mathcal P_{21,i}{\bm e_x}+\mathcal P_{22,i}{\bm e_y})=\frac{{\bm e_y}}{\sqrt{N}}\;,\\
 \bm p^{(2x)}_i&=\frac1{2\sqrt{N}}(\mathcal P_{11,i}{\bm e_x}-\mathcal P_{12,i}{\bm e_y}) \nonumber \\&=\frac1{\sqrt{N}}[\cos2\phi_i{\mathbf e_x}+\sin2\phi_i{\mathbf e_y}]\;,\\
\bm p^{(2y)}_i&=\frac1{2\sqrt{N}}(\mathcal P_{21,i}{\bm e_x}-\mathcal P_{22,i}{\bm e_y}) \nonumber \\&=\frac1{\sqrt{N}}[\sin2\phi_i{\mathbf e_x}-\cos2\phi_i{\mathbf e_y}]\;.
\end{align}\end{subequations}
It is then relatively straightforward to see that mode $1x$ only couples with the $x$-polarized wave and similarly the mode $1y$ only couples {with} the $y$-polarized {wave,} and the modes $2x$, $2y$ do not couple with the incident field at all; in that sense we can identify them to be dark modes.

At this point we would like to acknowledge that we have proven {{\em an important result}}: any response from a symmetric ring of particles that transforms according to $E$ (e.g. those excited by a normal incidence plane wave) will be spanned by explicitly {{\em four linearly independent basis vectors}}.  In regard to {eigenmodes,} this {means} that any such response will be described by no more than {{\em two doubly-degenerate eigenmodes}}, irrespective of how many particles the ring contains.  This is a generalization of previous results obtained for particular values of $N$~\cite{Bao2010}. Moreover, what we have shown here is that oligomers made of multiple rings contain two doubly-degenerate eigenmodes per a ring. This agrees with the existing studies of mechanical vibrations of the molecules. For instance, the benzene molecule has two, $D_{6h}$ rings made of  carbon and hydrogen atoms, and four (doubly-degenerate) in-plane vibration modes in total~\cite{wilson1934}. We can also consider the effect of a single central particle, the response of which will be described by a single dipole moment vector at the origin. Such an electric dipole moment must always transform according to $E$  (when neglecting the $z$-direction once again) and therefore the space of the associated responses will increase by two, which adds one more doubly-degenerate eigenmode.
So for a given polarization of the incident wave; the number of modes that can be excited in oligomers consisting of one ring without (with) a central particle is restricted to only two (three)  modes.  This is, counterintuitively, not dependent on the number of particles in the ring that make up the {nanoparticle} system.  
We stress here that the approach to now has {\em not} identified eigenmodes of the system.  
What is presented in Eq.~(\ref{eq:basis mode definitions}) was obtained based {solely} on the symmetry considerations and served to provide basis vectors for the response.   {Equation}~(\ref{eq:basis mode definitions}) is in fact a generalization of the existing understanding of modes in {oligomers, and it} shows that mode hybridization is required to excite the dark modes ($2x$ and $2y$). 
The opposing argument as to why these deduced dark modes can actually not exist in the real system of eigenmodes is then somewhat subtle.  The key is that the symmetry approach produces orthogonal modes by definition, whereas the interaction matrix {[see Eq.~(\ref{eq:alternate form}) in the next section]}, which describes the coupling between dipole moments of each particle,  is non-Hermitian~\cite{Merchiers2007}.  As {such,} the eigenmodes of the system are {\em not} orthogonal except in accidental cases.  This means that the space spanned by the dark basis vectors ($2x$ and $2y$) can be spanned by nonorthogonal bright vectors in the real eigenmode system.  
In this sense the projection operator approach has forced orthogonality onto the space of the system's responses and found dark basis vectors by necessity.  {Moreover,} an orthogonal basis which includes the incident field vectors ($1x$ and $1y$) requires that the remaining basis vectors be orthogonal to the incident field.  Admittedly, this does not explicitly rule out the existence of dark eigenmodes, which transform according to $E$, as we discuss in the following section.  
\section{Examples of nanoparticle oligomers}\label{sec:examples}
{In what follows, we present numerical simulations of the excitation of electric and magnetic eigenmodes in differnet plasmonic and all-dielectric oligomer structures to demonstrate the application of our theory.  Calculations are performed numerically using the coupled electric and magnetic dipole approximation~\cite{Mulholland1994}, where each constituent particle in an oligomer is modelled as a single electric and magnetic dipole.  A full description of this model, including the definition of the electric and magnetic eigenmodes, is provided in the next section.  Here we focus on the implications of the arguments provided in the previous section. To first}  illustrate the arguments on the number of eigenmodes that transform according to $E$ in a ring-type oligomer, we consider the most simple system that meets the associated symmetry requirements; {\em{a trimer}} with $D_{3h}$ symmetry.
\begin{figure}[h!]
\centering
\centerline{\includegraphics[width=0.95\columnwidth]{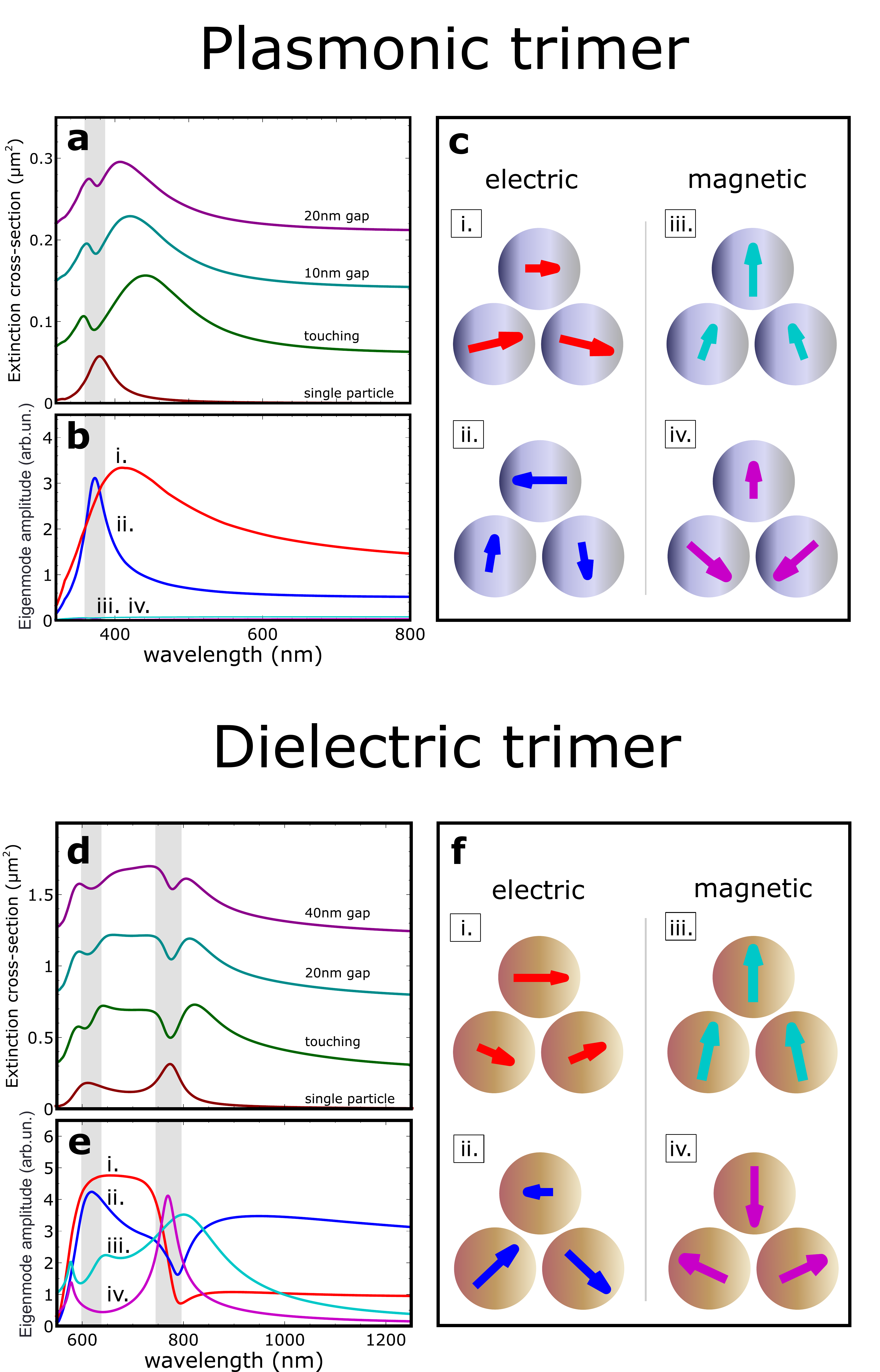}}
\caption{{Numerical} extinction spectra and associated eigenmode decompositions for (a-c) a silver nanosphere trimer and (d-f) a silicon nanosphere trimer as interparticle spacings are varied.    {Calculations are performed using the coupled electric and magnetic dipole approximation.}  The response of the trimers with 20~nm interparticle gaps are decomposed into electric and magnetic eigenmodes (b,e), where it can clearly be seen that the additional excitation of magnetic eigenmodes is non-negligible for the silicon trimer.  The real component of each eigenmode's dipole moment profiles are also shown (c,f) for electric and magnetic cases.    
In the extinction spectra for (a) silver  trimers and (d)  silicon trimers, a Fano lineshape occures at the intersection of electric and magnetic eigenmodes respectively.  Noticeably this, also corresponds to the single particle resonance. The silver nanospheres have a diameter of 80~nm and the silicon nanospheres a diameter of 200~nm.  Dispersion data is from Palik~\cite{Palik1998}.}
\label{fig:trimers}
\end{figure}
In Fig.~\ref{fig:trimers}, we show the {extinction spectra of plasmonic (silver) and dielectric (silicon) trimers} with their associated decomposition into both electric and magentic eigenmodes (the definition of electric and magnetic eigenmodes are provided in the following section).  In doing this decomposition, we notice that the magnetic modes have a negligible role in the {plasmonic} trimer, whereas {\em both electric and magnetic eigenmodes are excited in the {dielectric} trimer}.  In  both the cases, there are only ever two eigenmodes excited by the incident field for the electric and magnetic cases, which is in full agreement with {our} theory.  The reason why two, rather than four, eigenmodes are important for both electric and magnetic responses is that the polarization of the nondegenerate eigenmodes is defined to match the polarization of the incident field, and hence we only excite one eigenmode in each doubly-degenerate eigenspace.  

We start with {a} silver trimer, supporting a not-so-pronounced Fano resonance at the intersection of the two electric eigenmodes.  In other words, the Fano resonance occurring at the point where two eigenmodes have equal magnitude suggests that the interference of eigenmodes plays an important role, especially given the electric dipole moment profiles in Fig.~\ref{fig:trimers}(c) show that these eigenmodes are out of phase with each other.  Similarly, for the silicon trimer, there exists a well-pronounced Fano lineshape, but this time at the equivalent intersection of two magnetic eigenmodes.  Noticeably, the increase of the magnetic response also corresponds to a reduction in the overall electric response, which is due to the coupling between electric and magnetic dipoles.  Further intricacies of the dielectric system are apparent in that there is also a (smaller) secondary interference lineshape occurring at the peak of the sub-radiant electric eigenmode, which interferes with the super-radiant electric eigenmode.  That is to say there appears to be both electric and magnetic Fano-like interference occurring in one trimer.  

Additionally, it is also worth noting that there exists an apparent relationship between the location of the single-particle resonance and the location of the Fano resonance.  This is at least intuitive given that the sub-radiant modes (modes ii. and iv.) in {Figs}.~\ref{fig:trimers}(c,f)  have a comparatively low projection onto the incident field and subsequently their excitation is significantly dependent on interparticle coupling.  
As {such,} the single particle resonance represents a wavelength where the field radiating from each individual particle as a direct response to the incident field is maximised and it is therefore not surprising to see more significant excitation of the sub-radiant modes.

\begin{figure}[h!]
\centering
\centerline{\includegraphics[width=0.95\columnwidth]{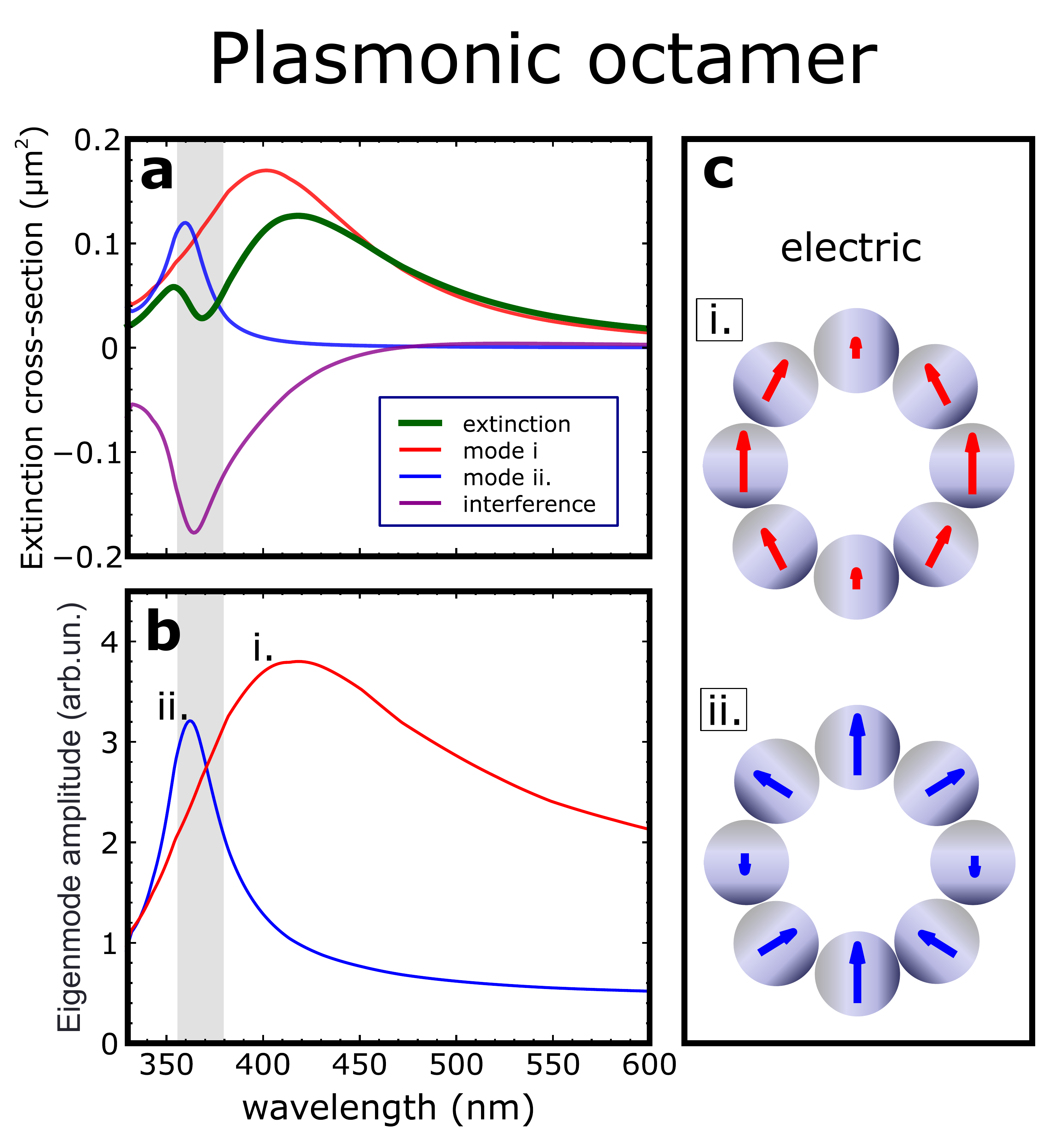}}
\caption{{Numerical simulation results for a silver nanosphere octamer under plane wave excitation showing the decomposition, in terms of the excitation of electric eigenmodes, of  (a) the extinction spectrum and (b) the total excitation of dipole moments.  The curves in the extinction decomposition correspond to the terms in Eq.~(\ref{eq:interference terms}).}  Also shown in (c) are the real components of each eigenmode's electric dipole moment profile.   Despite the larger number of particles, there remain only two nondegenerate eigenmodes in full agreement with the predictions of the theory provided in this section.  Additionally, much like the plasmonic trimer case in Fig.~\ref{fig:trimers}, the intersection of the two electric eigenmodes corresponds to a Fano resonance and the contribution of the mode interference to the extinction can explicitly be seen in (a) to lead to the creation of this Fano resonance.  The diameter of the silver nanospheres is 60~nm and consecutive nanospheres are touching in the octamer.  Silver dispersion data is from Palik~\cite{Palik1998}  {and the calculation of extinction spectra and eigenmodes is performed using the coupled electric and magnetic dipole approximation.}}
\label{fig:plasmonicOctamer}
\end{figure}
To demonstrate that the number of nondegenerate eigenmodes that can be excited by a normal-incidence plane wave is fixed independently on the number of particles in the given ring,  in Fig.~\ref{fig:plasmonicOctamer} we also consider a silver nanosphere octamer.  Here we do not include magnetic eigenmodes given the magnetic response is negligible, as was the case with the silver trimer in Fig.~\ref{fig:trimers}.   For the octamer it can be seen that the number of nondegenerate electric eigenmodes does remain fixed at precisely two despite the additional particles.

Similar to the trimer, we observe Fano resonances  in the extinction spectra of the octamer exactly at the intersection of two eigenmodes.   However, to fully justify the claim that the Fano resonance results from the interference of these eigenmodes, the extinction cross-section in Fig.~\ref{fig:plasmonicOctamer}(a) is decomposed into the components that come from each mode and also the component coming from the interference between them.  This interference term is a consequence of the eigenmodes not being orthogonal, a point which we can illustrate by considering an incident field written in terms of our two nondegenerate eigenmodes, 
\begin{subequations}
\begin{align}
&\left |{E_0} \right \rangle= a_1 \left | \nu _1 \right \rangle +  a_2\left |\nu _2 \right \rangle \\
&\Rightarrow \left |p\right \rangle = a_1 \lambda_1 \left | \nu _1 \right \rangle +  a_2 \lambda_2  \left |\nu _2 \right \rangle\;.
\end{align}
\end{subequations}
{ Here the coefficients of each eigenmode in the induced dipole moment solution (i.e. the $a_i \lambda_i$ terms) are the eigenmode amplitudes that are plotted in figures; they represent the magnitude of excitation for each eigenmode.  However the extinction cross-section~\cite{Draine:94} is proportional to}
\begin{subequations}\begin{align}
\sigma_e \propto & \mathrm{Im}\left \{ \left \langle {E_0}  |{p} \right \rangle \right \} \;,\label{eq:extinction zero reference}\\
\left \langle {E_0}  |{p} \right \rangle   =& (a_1^* \left \langle {\nu_1} \right |  +  a_2^* \left \langle {\nu_2} \right | )
(a_1 \lambda_1 \left |\nu _1 \right \rangle +  a_2 \lambda_2 \left |\nu _2 \right \rangle)\\
=&  \underset {\mathrm{direct\;terms}} {\underbrace{|a_1|^2\lambda_1  | \nu_1|^2 + |a_2|^2 \lambda_2  | \nu_2|^2    }} \nonumber \\
&  + \underset {\mathrm{interference\;terms}} {\underbrace{   a_1^* a_2  \lambda_2 \left \langle {\nu _1}  |{\nu _2} \right \rangle 
 + a_2^* a_1  \lambda_1\left \langle {\nu _2}  |{\nu _1} \right \rangle }}\;. \label{eq:interference terms}
\end{align} \end{subequations}
In this expression, the ``direct terms'' correspond to the contribution of each eigenmode to the extinction without accounting for the interplay between eigenmodes (the ``interference terms").  {It is precisely the direct terms for each eigenmode and the sum of interference terms seen in Eq.~(\ref{eq:interference terms}) that are presented in Fig. \ref{fig:plasmonicOctamer}(a) for the silver nanosphere octamer.}  It is important to note that each of the direct terms can only  produce a positive contribution to the exitinction as they represent the extinction in the case where the incident field has been structured to consist of only the one given eigenmode.  Indeed, if this was not the case the energy conservation law would be violated. 
However, as seen in Fig.{\ref{fig:plasmonicOctamer}}(a), the interference terms can lead to a negative contribution to extinction. Such a negative contribution has been observed previously~\cite{Frimmer2012}, however it is important  that it is the interference alone that can produce a negative extinction component.  In other words, no eigenmode can feed energy into the incident field.  It is also worth noting that if we were to have an orthogonal basis of eigenmodes the interference terms all go to zero as is the case with analogous inner products in quantum mechanics and other Hermitian spaces.  In this sense, we can conclude that the Fano lineshape is not a result of  energy coupling into a non-radiative dark mode, but rather the interference between the two eigenmodes.  
Additionally, a further effect of a nonorthogonal basis of eigenmodes is that the decomposition of the incident field into eigenmodes has to be solved through a set of coupled equations.  Specifically if we were to consider the decomposition of some incident field in terms of eigenmodes
\begin{align}
\left |{E_0} \right \rangle= \sum \limits_i a_i \left | \nu _i \right \rangle \:,
\end{align}
such a decomposition, $\{ a_i\}$, would then have to be solved through a set of equations such as
\begin{align}
\left \langle \nu _j |{E_0} \right \rangle= \sum \limits_i a_i \left \langle \nu _j | \nu _i \right \rangle \quad \forall \; j = 1,\;...\;N\;. \label{eq:coupled projections}
\end{align}
In this way, it is worth noting that even if the eigenmodes are themselves decoupled; their excitations {\em are} coupled.   In regard to the possibility of dark eigenmodes; Eq.~(\ref{eq:coupled projections}) means that an eigenmode can be excited even if it has a zero-projection onto the incident field (i.e. if it is dark).  In this case we can actually use the derived basis vectors in Eq.~(\ref{eq:basis mode definitions}) to get an explicit expression for the dark eigenmode of the arbitrary system.  
Moreover, because an arbitrary eigenmode of a ring-type oligomer can be written as a linear combination of the set of {\em orthogonal} basis vectors provided in Eq.~(\ref{eq:basis mode definitions}); a dark eigenmode having a zero projection onto both $1x$ and $1y$ (i.e. being also the x- and y-polarized incident field) ensures that it is expressible entirely in terms of the remaining two basis vectors: $2x$ and $2y$.  Then, given that such an eigenmode is doubly-degenerate because it transforms according to $E$, we know it must have a two-dimensional eigenspace and that there will subsequently be two orthogonal eigenmodes that span this dark eigenspace.  Said another way; if there is a dark eigenmode then the basis vectors $2x$ and $2y$ from Eq.~(\ref{eq:basis mode definitions}) will span precisely the associated dark eigenspace and therefore these two basis vectors will be the dark eigenmodes of a ring-type oligomer.  However, it is nonetheless worth noting that the existence of dark modes that transform according to $E$ is not guaranteed in any oligomer and it is unlikely that they would be frequency-independent as $2x$ and $2y$ are.  Thus, dark eigenmodes would be isolated  much like the pure super-radiant eigenmodes (i.e. making the incident field an eigenmode), which are the other  basis vectors presented in Eq.~(\ref{eq:basis mode definitions}).  In this regard, all numerically-calculated eigenmodes in this paper are explicitly bright at all calculated frequencies.

\begin{figure}[h!]
\centering
\centerline{\includegraphics[width=0.95\columnwidth]{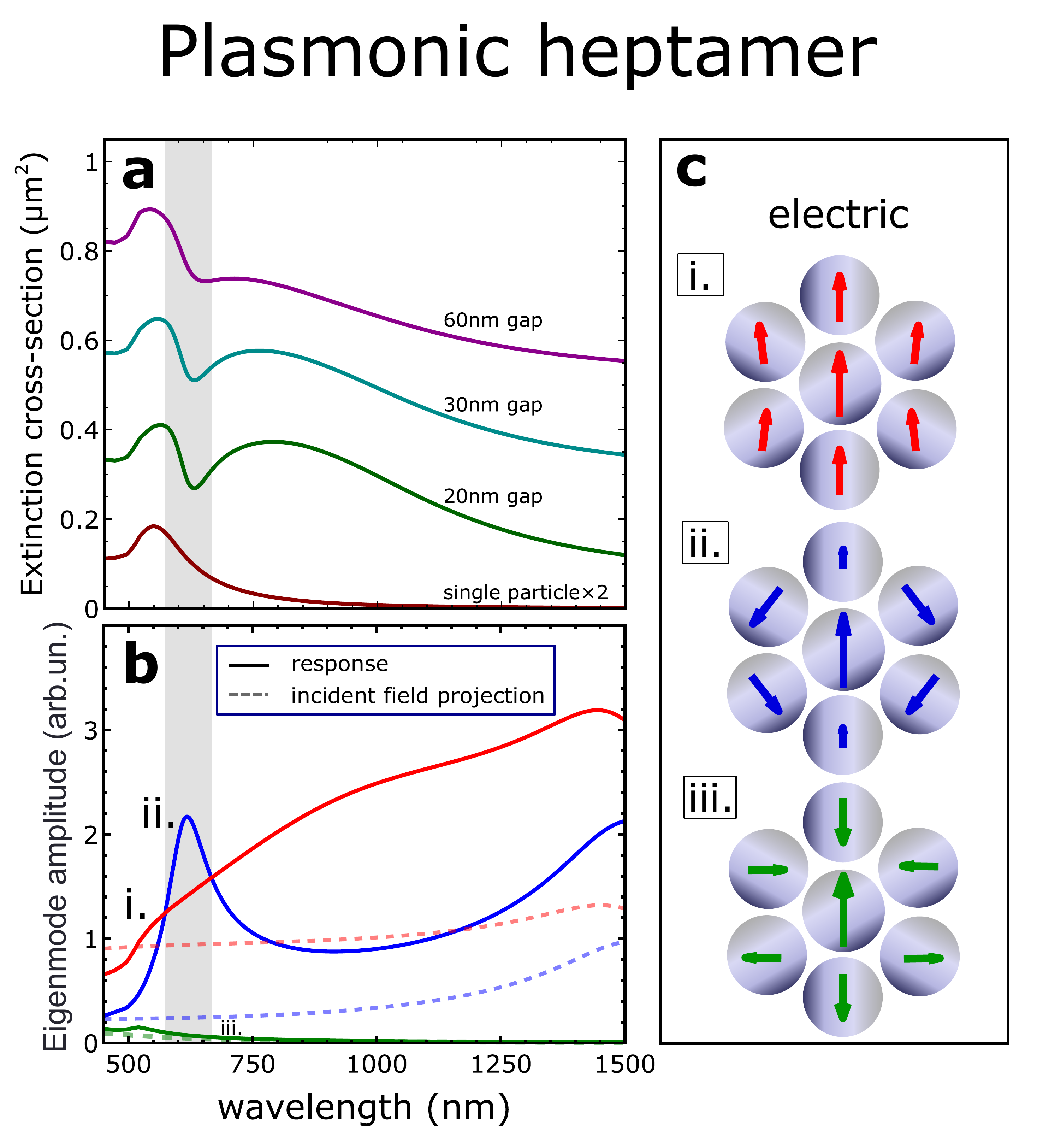}}
\caption{(a) {Numerically simulated} extinction spectra of a gold nanosphere heptamer with varying interparticle gaps, (b) the associated decomposition of the heptamer's response into three nondegenerate eigenmodes for the 30~nm gap case and (c) the real components of each eigenmode's dipole moment profiles at the Fano resonance frequency. In (b) the decomposition of the heptamer's response into eigenmodes (unbroken lines) shows that the Fano resonance is being created when one sub-radiant eigenmode (ii.) experiences a narrow resonance over a broad resonance of the super-radiant mode (i.).  Further each of these eigenmodes has a nonzero projection onto the incident field and they are therefore bright.  All gold nanospheres have a diameter of 150~nm and the gold dispersion data comes from Palik~\cite{Palik1998}. {The calculation of extinction spectra and eigenmodes is performed using the coupled electric and magnetic dipole approximation.}}
\label{fig:plasmonicHeptamer}
\end{figure}

To  demonstrate the applicability of this eigenmode theory to experiment, we reconsider the gold heptamer made by Hentschel et al.~\cite{Hentschel2010}  in Fig.~\ref{fig:plasmonicHeptamer}.  This investigation observed a Fano resonance experimentally and associated it with coupling into a dark mode.   
However, based on our analysis one can explicitly see that such Fano resonances are described by the inference of two bright eigenmodes. Moreover Fig.~\ref{fig:plasmonicHeptamer}(b) shows the eigenmode decomposition of the system's response (unbroken line), but also the projections of each eigenmode onto the incident electric field (dashed line), which are all nonzero.  

This is not the same as the previously {supported} explanation in a number of papers, which attributed the Fano resonance to the coupling of a bright mode {with} a dark mode.  It is worth noting that the bright modes from our simulation appear to have dipole moment profiles that qualitatively match that of the modes and dark modes proposed by Fan et al.~\cite{Fan2010} for a similar gold ring heptamer structure.  The key point of our analysis applied to this structure is therefore not that the mode profiles are largely different, but rather that the true modes are bright instead of dark.  
{However,} we can conclude that the Fano resonance in this system is explicitly due to the interference between bright modes only, and {it} does not involve {any} dark mode as previously {agreed}.

\section{Dielectric oligomers} 
\label{sec:dielectrics}
{The electric dipole response is known to be the dominant contribution to the optical properties of a single spherical particle as its size is made arbitrarily small, so called Rayleigh approximation. This is the principle on which the coupled electric dipole approximation is based~\cite{Draine:94} .  However, for high-refractive index dielectric materials such situation can change.  In particular, when the wavelength inside a particle matches its diameter a magnetic dipole mode can be also excited due to resonant excitation of the circulating displacement current.  In this way it is possible for the magnetic dipole response of a dielectric particle to be on the same order of magnitude as the electric dipole response~\cite{Mulholland1994}.  Importantly, provided the particle also remains small compared to the wavelength, it will also be substantially larger than its electric quadrupole response, which is otherwise the next dominant order of multipole after the electric dipole response in plasmonic materials~\cite{Evlyukhin2012}.  
Therefore, in such a regime, it is necessary to take into account the magnetic dipole responses for high-refractive index dielectric materials and, subsequently, an all-dielectric oligomer requires full treatment of both electric and magnetic dipole moments to characterize its behavior. 
In the following, we show that the electric and magnetic responses of symmetric oligomers can be considered independently with distinct electric and magnetic eigenmodes as opposed to combined electromagnetic eigenmodes.} 

To begin, the general expressions for the electric and magnetic dipole moments of the $i^{\mathrm{th}}$ particle in an oligomer can be defined using the coupled electric and magnetic dipole equations~\cite{Mulholland1994}
\begin{subequations}\label{eq:original dipole equations}
\begin{align}
\mathbf{p}_{i}  = \alpha_{E,i}\epsilon_{0}\mathbf{E_{0}}(\mathbf{r}_{i})&+\alpha_{E,i}\epsilon_{0}k^{2}
  \left(\underset{i\neq j}{\overset{N}{\sum}}\frac{1}{\epsilon_{0}}\hat{G}^{0}(\mathbf{r}_{i},\mathbf{r}_{j}) \mathbf{p}_{j} \right.   \nonumber  \\ & \left.  -\sqrt{\frac{\mu_{0}}{\epsilon_{0}}}\nabla\times\hat{G}^{0}(\mathbf{r}_{i},\mathbf{r}_{j}) \mathbf{m}_{j}\right) ,\\
 \mathbf{m}_{i} =  \alpha_{H,i}\mathbf{H_{0}}(\mathbf{r}_{i})&  +\alpha_{H,i}k^{2}  \left(\underset{i\neq j}{\overset{N}{\sum}}\hat{G}^{0}(\mathbf{r}_{i},\mathbf{r}_{j}) \mathbf{m}_{j}\right.   \nonumber  \\ & \left. +\frac{1}{\sqrt{\epsilon_{0}\mu_{0}}}\nabla\times\hat{G}^{0}(\mathbf{r}_{i},\mathbf{r}_{j}) \mathbf{p}_{j}\right),
\end{align}
\end{subequations}
where $\mathbf{p}_i$ ($\mathbf{m}_i$) is the electric (magnetic) dipole moment of the $i^{\mathrm{th}}$ particle, $\hat{G}^{0}(\mathbf{r}_{i},\mathbf{r}_{j}) $ is the free space dyadic Greens function between the $i^{\mathrm{th}}$ and $j^{\mathrm{th}}$ particle, $\alpha_{E,i}$ ($\alpha_{H,i}$) is the electric (magnetic) polarizability of the $i^{\mathrm{th}}$  particle and $k$ is the wave number of light.  The free space dyadic Green's functions are defined as
\begin{subequations}\label{eq:Cij and fij definitions}
{\small \begin{align}
\hat{G}^{0}(\mathbf{r}_{i},\mathbf{r}_{j}) & =  \hat{G}_{ij}=a_{ij}\hat{I}+b_{ij}\mathbf{n}_{ji}\otimes\mathbf{n}_{ji} \;,\\
\nabla\times\hat{G}^{0}(\mathbf{r}_{i},\mathbf{r}_{j}) & =  \hat{g}_{ij}=d_{ij}\left(\mathbf{n}_{ji}\otimes\mathbf{n}_{ji}-\hat{I}\right)^{\frac{1}{2}}\;,
\end{align}}
\end{subequations}
where $\hat I$ is the identity matrix, $\mathbf{n}_{ji}$ is the unit vector pointing from the $j^{\mathrm{th}}$ to $i^{\mathrm{th}}$ particle, the  $\mathbf{n}_{ji}\times\hat{I}^{(3)}$ operator is expressed explicitly as
$\left(\mathbf{n}_{ji}\otimes\mathbf{n}_{ji}-\hat{I}^{(3)}\right)^{\frac{1}{2}}$ and the following scalars have been defined
\begin{subequations}\label{eq:CDE coefficients}
{\small \begin{align}
a_{ij} & =  { \frac{e^{ik r_{ji}}}{4\pi r_{ji}}\left(1+\frac{i}{k r_{ji}}-\frac{1}{k^{2}r_{ji}^{2}}\right)} \;,\\
b_{ij} & =  { \frac{e^{ik r_{ji}}}{4\pi r_{ji}}\left(-1-\frac{3i}{k r_{ji}}+\frac{3}{k^{2}r_{ji}^{2}}\right)} \;,\\
d_{ij} & =  { \frac{e^{ik r_{ji}}}{4\pi r_{ji}}\left(1+\frac{i}{k r_{ji}}\right)}\;,
\end{align}}
\end{subequations}
for $r_{ji}=|\mathbf{r}_{i}-\mathbf{r}_{j}|$.\\
{In all simulations presented in this paper the dipole polarizabilities are calculated for spheres using the Mie theory dipole scattering coefficients $a_{1},\: b_{1}$
\begin{subequations}
 \begin{align}
\alpha_{E} & =  6i\pi a_{1}/k^{3}\;, \\
\alpha_{H} & =  6i\pi b_{1}/k^{3} \;.
\end{align}
\end{subequations}}
The coupled-dipole equations in Eq.~(\ref{eq:original dipole equations}) can then be written in an equivalent matrix notation
\begin{subequations} \label{eq:dipole equations 1}
\begin{align}
 \epsilon_0  \left|E_{0}\right\rangle  = \left(  \hat{\alpha}_E^{-1} - k^{2} \hat{M}_G\right)\left|p\right\rangle  + \sqrt{\epsilon_0 \mu_0}  k^{2} \hat{M}_g\left|m\right\rangle \;, \\
 \left|H_{0}\right\rangle  = \left(  \hat{\alpha}_H^{-1} -  k^{2} \hat{M}_G\right)\left|m\right\rangle   -  \frac{k^{2}}{\sqrt{\epsilon_0 \mu_0}} \hat{M}_g\left|p\right\rangle \;,
\end{align}
\end{subequations}
where $\hat{\alpha}_{E}$ ($\hat{\alpha}_{H}$) is a diagonal matrix containing the electric (magnetic) polarizabilities of each particle and $\hat{M}_{G}$ or $\hat{M}_{g}$ contains the appropriate collection of dyadic Green's functions to act between the dipoles	
\begin{subequations}	
\begin{align} \label{eq:Mg and MG definitions} 
\hat{M}_G & =  \left(\begin{array}{cccc}
\hat{0} & \hat{G}_{12} & \cdots & \hat{G}_{1N}\\
\hat{G}_{12}& \hat{0} &  & \vdots\\
\vdots &  & \ddots\\
\hat{G}_{1N} & \cdots &  & \hat{0}
\end{array}\right)\;,\\
 \hat{M}_{g}  &=  \left(\begin{array}{cccc}
\hat{0}& -\hat{g}_{12} & \cdots & -\hat{g}_{1N}\\
\hat{g}_{12}& \hat{0} &  & \vdots\\
\vdots &  & \ddots\\
\hat{g}_{1N}& \cdots &  & \hat{0}
\end{array}\right) \;.
\end{align}
\end{subequations}	

Given the matrix equations seen in Eq.~(\ref{eq:dipole equations 1}), it is apparent that there is flexibility in whether the incident field is written as a function of electric and magnetic dipole moments, or whether the electric or magnetic dipole moments are written as a function of the incident electric and magnetic fields.  Moreover, by rearranging, we can write Eq.~(\ref{eq:dipole equations 1}) in the following alternate form
\begin{subequations} \label{eq:alternate form}
\begin{align}
\left|p\right\rangle &=   \epsilon_0\hat{M}_{ee} \left ( \left|E_{0}\right\rangle - \sqrt{\frac{ \mu_0} {\epsilon_0}}  \hat{M}_{eh} \left|H_{0}\right\rangle   \right)\;,   \\
 \left|m\right\rangle & = \hat{M}_{hh} \left(  \left|H_{0}\right\rangle + \sqrt{\frac{\epsilon_0}{\mu_0}}\hat{M}_{he} \left|E_{0}\right\rangle   \right)\;,
\end{align}
\end{subequations}
where the $\hat{M}$ matrices have been defined as
\begin{subequations} 
\begin{align}
 \hat{M}_{ee} &=  \left( \hat{\alpha}_{E}^{-1} - k^{2} \hat{M}_G +k^4 \hat{M}_g \left( \hat{\alpha}_{H}^{-1} - k^2 \hat{M}_G \right)^{-1}\hat{M}_g \right) ^{-1} \;, \\
 \hat{M}_{hh} &=\left( \hat{\alpha}_{H}^{-1} - k^{2} \hat{M}_G +k^4 \hat{M}_g \left( \hat{\alpha}_{E}^{-1} - k^2 \hat{M}_G \right)^{-1}\hat{M}_g \right) ^{-1}  \;,\\
\hat{M}_{eh} &=  k^{2} \hat{M}_g \left( \hat{\alpha}_{H} ^{-1} -  k^2 \hat{M}_G \right)^{-1} \;, \\
\hat{M}_{he} &=  k^{2} \hat{M}_g \left( \hat{\alpha}_{E} ^{-1} -  k^2 \hat{M}_G \right)^{-1}   \;.
\end{align}
\end{subequations}
Here the matrices $\hat{M}_{ee}$ and  $\hat{M}_{hh}$ describe the response of the system to applied forcings whereas  $\hat{M}_{eh}$ and  $\hat{M}_{he}$ describe the component of bi-anisotropic forcing on the electric and magnetic dipoles that comes from the magnetic and electric fields respectively (both $\hat{M}_{eh}$ and  $\hat{M}_{he}$ are singular).  The obvious change in using the form of the coupled-dipole equations in Eq.~(\ref{eq:alternate form}) is that the electric and magnetic dipole moments can be calculated separately to each other.  That is to say, rather than using one rank $6N$ matrix to describe the electromagnetic system~\cite{Merchiers2007} we are able to instead use the two rank $3N$ matrices ($\hat{M}_{ee}$ and  $\hat{M}_{hh}$) to describe how the system responds electrically and magnetically (respectively).  The reason this particular formulation is important is because the electric and magnetic forcings are distinct from each other and hence each forcing transforms according to a single irreducible representation (when considering a normal incidence plane wave).  This therefore makes the electromagnetic dipole system directly subject to the eigenmode arguments provided in the previous section.  It is important to note that no additional approximations have been made in writing the dipole equations in the manner of Eq.~(\ref{eq:alternate form}) - this formulation still represents the full electromagnetic dipole system.

Now, for the specific case of oligomers with $D_{Nh}$ symmetry, we are able to restrict the observed effect of the bi-anisotropic forcings.  
This is specifically because, under normal plane wave excitation, the incident field is acting uniformly on all dipoles in planar systems and therefore it is only the sum of dipole moments in the complete oligomer, the `total dipole moment', that affects the extinction cross-section. Moreover the extinction cross-section can be written as
\begin{subequations} \label{eq:extinction expression}
{\small \begin{align}
\sigma_{e} & =  \frac{k}{\epsilon_{0}\left|\mathbf{E}^{0}\right|^{2}}\mathrm{Im}\left\{ \sum\limits_{i} \mathbf{E_0}^{\dagger}\mathbf{p}_{i} + \mathbf{H_0}^{\dagger}\mathbf{m}_{i} \right\} \;,\\
& = \frac{k}{\epsilon_{0}\left|\mathbf{E}^{0}\right|^{2}}\mathrm{Im}\left\{  \mathbf{E_0}^{\dagger}\left( \sum\limits_{i}\mathbf{p}_{i} \right) + \mathbf{H_0}^{\dagger} \left(\sum\limits_{i} \mathbf{m}_{i} \right) \right\} \;.
\end{align}}
\end{subequations}
In a physical sense this means that only the total dipole moment of such a structure can couple directly to an arbitrarily-polarized incident field.  
So,by  using Eq.~(\ref{eq:alternate form}), the total dipole moments can be written as a function of the electric and magnetic incident field vectors
\begin{subequations} \label{eq:total dipole moments1}
\begin{align}
\sum \limits_i \mathbf{p}_i &=  \epsilon_0 \hat{\mathcal{M}}_{ee} \mathbf{E_0}(\mathbf{\mathbf{r}_0})   -\sqrt{\epsilon_0 \mu_0} \;  \hat{\mathcal{M}}_{eh} \mathbf{H_0}(\mathbf{\mathbf{r}_0})\;,   \\
\sum \limits_i \mathbf{m}_i &=  \hat{\mathcal{M}}_{hh} \mathbf{H_0}(\mathbf{\mathbf{r}_0})  +\sqrt{\frac{\epsilon_0}{\mu_0}}  \;\hat{\mathcal{M}}_{he} \mathbf{E_0}(\mathbf{\mathbf{r}_0})\;,
\end{align}
\end{subequations}
where we have defined the $3\times3$   matrices, $\hat{\mathcal{M}}$, as
\begin{align}
 \hat{\mathcal{M}}_{ee,hh, eh,he} &= \underset{3\times3\;\mathrm{blocks}} {\sum} \hat{M}_{ee,hh, eh,he}\;. \label{eq:total dipole moments2}  
\end{align}

However, in a structure with $D_{Nh}$ symmetry, the $\hat{G}_{ij}$ and $\hat{g}_{ij}$ matrices react oppositely (even or odd)  when applying the in-plane reflection symmetry operation, $\sigma_{h}$, to the position vectors.  Specifically the following relation holds 
\begin{subequations} \label{eq:this is good}
\begin{align}
\hat G_{ij} &= \hat\sigma_{h}\hat G_{ij} \hat\sigma_{h}^{\dagger}\;, \\
\hat g_{ij} &= -\hat\sigma_{h}\hat g_{ij}\hat{\sigma_{h}}^{\dagger} \;,
\end{align}
\end{subequations}
where $\hat{\sigma}_{h}$ is defined as the local reflection operation that acts on a position vector
\begin{align}
\hat{\sigma}_{h} &= \left( \begin{array}{ccc} 1  &0 &0\\ 0& 1&0 \\0 &0 &-1
\end{array}\right) \;.
\end{align}
However, given $D_{Nh}$ contains a $C_N$ symmetry subgroup, it has previously been shown~\cite{Hopkins2013} that we will have forms for the summed $\hat{\mathcal{M}}$ matrices that always commute with  $\hat\sigma_{h}$
\begin{align}\hat{\mathcal{M}}&\;\propto\; \left( \begin{array}{ccc}
a & c & 0  \\ -c & a & 0  \\0 &0 & b
\end{array}\right) \qquad a,b,c \in \mathbb{C}  \;.\label{eq:Cn form}
\end{align}
For this reason, applying the reflection symmetry operation to position vectors in Eq.~(\ref{eq:total dipole moments2}) in the manner described by Eq.~(\ref{eq:this is good}) will require that both the aggregate electric to magnetic coupling matrices ($\hat{\mathcal{M}}_{eh}$ and $\hat{\mathcal{M}}_{he}$) be equal to their own negative.  In other words these matrices must be zero and subsequently the bi-anisotroptic forcing will have no effect on the total dipole moments or extinction cross-section.  This can understood from the perspective that $D_{Nh}$ symmetry is not chiral, which is typically the catalyst that leads to such bi-anisotropy.  
So, for a planar oligomer with $D_{Nh}$ symmetry, the dipole moments of the full electromagnetic system {[Eq.~(\ref{eq:total dipole moments1})]} will reduce to
\begin{subequations}
\begin{align}
 \sum \limits_i \mathbf{p}_i &= \epsilon_0 \hat{\mathcal{M}}_{ee}  \mathbf{E_0}(\mathbf{\mathbf{r}_0})\;, \\
\sum \limits_i \mathbf{m}_i &=  \hat{\mathcal{M}}_{hh} \mathbf{H_0}(\mathbf{\mathbf{r}_0}) \;.
\end{align}
\end{subequations}
As such, when considering far-field measures such as extinction [Eq.~(\ref{eq:extinction expression})] we are able to use the following, modified, dipole equations with no loss of generality or accuracy.  
\begin{subequations} \label{eq: reduced alternate form}
\begin{align}
\left|p\right\rangle &=  \epsilon_0 \hat{M}_{ee} \left|E_{0}\right\rangle \;, \\
 \left|m\right\rangle & = \hat{M}_{hh} \left|H_{0}\right\rangle \;.
\end{align}
\end{subequations}
In this way we can use the two matrices, $\hat{M}_{ee}$ and $\hat{M}_{hh}$, to define and calculate separate electric and magnetic eigenmodes rather than combined electromagnetic eigenmodes.  {For all simulations presented in this paper, we calculate the eigenmodes numerically from specifically these two matrix definitions.}

\begin{figure}[!ht]
\centering
\centerline{\includegraphics[width=0.95\columnwidth]{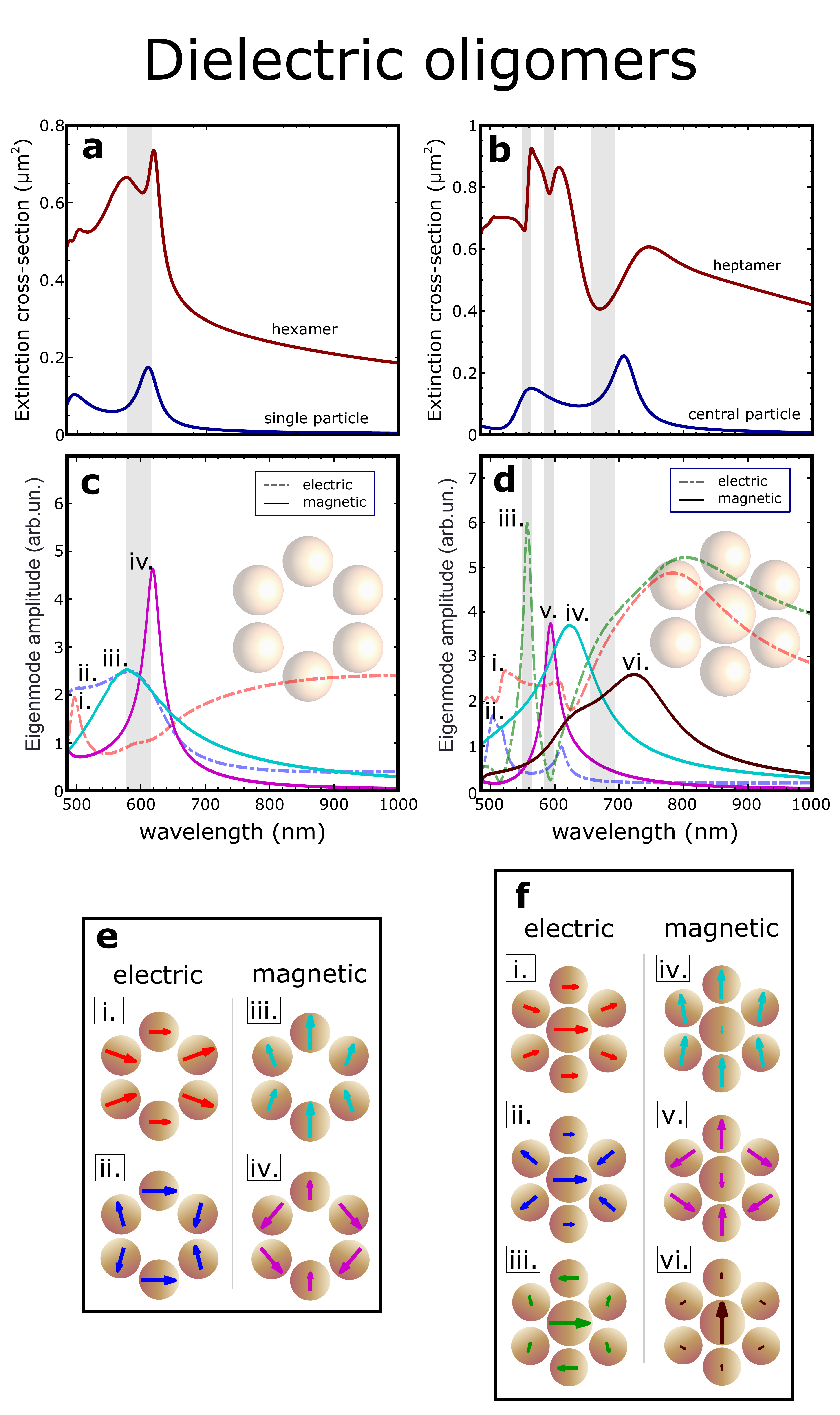}}
\caption{{Numerical} extinction spectra for the (a) surrounding hexamer (no central particle)  and (b) silicon nanosphere heptamer.   The eigenmode decomposition of the electric and magnetic responses is shown for the  (c) hexamer and (d) heptamer along with (e,f) the associated electric and magnetic dipole profiles at 700nm (real components).  Multiple Fano resonances can be seen occurring as a result of sharp resonances of eigenmodes and intersections of resonant eigenmodes.    The central nanosphere has a diameter of 180nm and the nanospheres in the surrounding hexamer have diameters of 150nm {being} separated with gaps of 20nm. The hexamer and heptamer extinction spectra have been shifted upwards for clarity.  Silicon dispersion data is from Palik~\cite{Palik1998}.  }
\label{fig:dielectricHeptamer}
\end{figure}

To demonstrate how Fano resonances can occur in all-dielectric oligomers we will now consider the silicon heptamer seen in Fig.~\ref{fig:dielectricHeptamer}.  Here there are three magnetic eigenmodes and three electric eigenmodes that interfere with each other to produce multiple Fano resonances and other sharp features.  Firstly, in regard to the analysis of all-dielectric oligomer Fano {resonances~\cite{Miroshnichenko2012}}, we can see that the resonance of the central particle corresponds to the largest Fano resonance.  This relationship can be reconciled with the eigenmode decomposition presented here given the dominance of the central particle in the eigenmode denoted by vi. in Fig.~\ref{fig:dielectricHeptamer}(f).  However there are also two other Fano resonances resulting from the sharp resonances of eigenmodes iii. (electric) and v. (magnetic)  as seen in Fig.~\ref{fig:dielectricHeptamer}(d), which do not correspond to the central particle resonance.   
When removing the central particle only one of these Fano resonances remains with only a slight frequency shift and broadening as seen in Fig.~\ref{fig:dielectricHeptamer}(a).  This can be expected given that the interfering (magnetic) eigenmode profiles look similar in both heptamer and hexamer cases when neglecting the central particle.  That is to say this Fano resonance is coming primarily from the ring-type oligomer in both cases.  Fano resonances in ring-type oligomers have been demonstrated earlier in this paper so here we will instead highlight the more subtle effect that adding/removing the central particle has on this one resonance.  
Specifically the shift and narrowing of this particular Fano resonance, from adding the central particle, can be seen to be a result of a narrower excitation of the corresponding eigenmode, but it is also worth noting that the mode itself becomes more subradiant in the heptamer case given the central particle oscillates out of phase with the surrounding hexamer.  In this sense the resonance in the heptamer closer resembles a `definition Fano resonance' where a narrow sub-radiant resonance interferes with a broad super-radiant mode.  This sub-radiance and increased similarity to textbook Fano resonance conditions correspondingly results in the Fano resonance occurring at the peak of the sub-radiant mode in the heptamer rather than the intersection between modes in the hexamer.  As such we are able to differentiate interference between two resonant eigenmodes and  these definition Fano resonances, although noticeably both lead to similar lineshapes and sharp features in extinction so the necessity for such a distinction is not particularly obvious.  
However the resulting resonance observed in extinction is not always simple given systems such as the heptamer can have more than two interfering eigenmodes.  For instance we observe that a definition Fano resonance is occurring from the interference of super- and sub-radiant electric eigenmodes (i. and iii. respectively) in Fig.~\ref{fig:dielectricHeptamer}(b), but the Fano lineshape becomes skewed by the third electric eigenmode. 
Understandably more thorough analysis can be done on this system.  However what we have aimed to demonstrate here is that the approach presented in this paper offers a richer framework to understand the variety of electromagnetic responses and their interactions in all-dielectric oligomers.

\section{Conclusions}

We have rigorously shown that there exists a common {physical} mechanism  leading to  Fano resonances in both plasmonic and all-dielectric symmetric oligomers.  The observed Fano lineshapes are attributed to the fact that the eigenmodes of oligomers are not orthogonal and, therefore, that they can interfere with each other.  We have also proved a key theorem to demonstrate that there {exists} a fixed number of eigenmodes that can be excited by a plane wave under normal incidence in a symmetric oligomer, and that this number is independent of the number of nanoparticles in the oligomer.  As such, we have shown that Fano resonances  can be realised in completely symmetric oligomers excited by normal incidence plane waves without involving additional complexity into the system.   
The presented approach  has been directed at addressing the need for analytical background in the rapidly growing field of all-dielectric systems and oligomers.  In addition, we have also  demonstrated that Fano resonances do not require the excitation of dark modes.  The mechanism presented here therefore also has an impact on the current surging interest into the study of the far-field effects of dark modes,~\cite{Gomez2013,Rahmani2011, Liu2013} because it facilitates similar effects through a simpler mechanism.
Finally, it is also worth noting that the approach presented in this paper is sufficiently general to be applicable to broader fields such molecular optics, antenna design and other applications that are able to utilize dipole-dipole interactions.  

{ After the submission of this manuscript a related study has been published by Forestiere et al.~\cite{Forestiere2013}. The authors employed a very different approach, but  they came to the same conclusion that Fano resonances in plasmonic oligomers are induced by the interference of collective eigenmodes without requiring the excitation of dark modes.}

\section{Acknowledgements}

The work of A.N.P. was supported by Ministry of Education and Science of the Russian Federation (Grants No. 11.G34.31.0020,0, 11.519.11.2037, 14.B37.21.0307,01201259765),  the Russian Foundation for Basic Research, European Union (projects POLAPHEN and SPANLG4Q). A.E.M. would like to acknowledge the financial support from the Australian Research Council via Future Fellowship program (FT110100037).  

\bibliographystyle{apsrev4-1}
\bibliography{bibliography}

\end{document}